\documentclass{emulateapj}
\usepackage{natbib}

\newcommand{\mjup}{$\cal{M}_{\rm Jup}$}

\newcommand{\ldl}{$\lambda/{\Delta}{\lambda}$}
\newcommand{\logg}{$\log{g}$}
\newcommand{\teff}{$T_{\rm eff}$}

\newcommand{\meth}{CH$_4$}
\newcommand{\amo}{NH$_3$}
\newcommand{\wat}{H$_2$O}

\newcommand{\kms}{km~s$^{-1}$}
\newcommand{\cmss}{cm~s$^{-2}$}
\newcommand{\name}{UGPS~J072227.51$-$054031.2}
\newcommand{\namesh}{UGPS~0722$-$05}
\newcommand{\vsini}{$v\sin i$}

\shortauthors{Bochanski et al.}
\shorttitle{FIRE Observations of {\namesh}}
\begin{document}

\title{FIRE Spectroscopy of the ultra-cool brown
  dwarf, {\name}:  Kinematics, Rotation and Atmospheric Parameters\footnote{This paper includes data gathered with the 6.5 meter Magellan Telescopes located at Las Campanas Observatory, Chile.}}

\slugcomment{Accepted by AJ}

\author{John J. Bochanski\altaffilmark{2,3},	
  Adam J. Burgasser\altaffilmark{3,4,5},                  
  Robert A. Simcoe\altaffilmark{3,6}, 
  Andrew A. West\altaffilmark{7,8} }
\altaffiltext{2}{Astronomy and Astrophysics Department, Pennsylvania
  State University, 525 Davey Laboratory, University Park, PA 16802, USA\\ 
email:jjb29@psu.edu}
\altaffiltext{3}{Kavli Institute for Astrophysics and Space Research,
  Massachusetts Institute of Technology, Building 37, 77 Massachusetts
  Avenue, Cambridge, MA 02139, USA}
\altaffiltext{4}{Center for Astrophysics and Space Science, University
of California San Diego, La Jolla, CA 92093, USA}
\altaffiltext{5}{Hellman Fellow}
\altaffiltext{6}{Sloan Fellow}
\altaffiltext{7}{Department of Astronomy, Boston University, 725
  Commonwealth Avenue, Boston, MA 02215, USA}
\altaffiltext{8}{Visiting Investigator, Department of Terrestrial Magnetism, Carnegie
Institution of Washington, 5241 Broad Branch Road, NW, Washington, DC
20015, USA}

\begin{abstract}
We present {\ldl} $\sim$ 6000 near--infrared spectroscopy of the nearby T9 dwarf, {\name}, obtained during the
commissioning of the Folded--Port Infrared Echellette Spectrograph on the Baade Magellan telescope at Las Campanas Observatory.  The spectrum is marked by
significant absorption from {\wat}, {\meth} and H$_2$.  We also identify {\amo}
absorption features by comparing the spectrum to recently published line lists.  The
spectrum is fit with BT-Settl models,
indicating {\teff} $\sim 500-600$ K and {\logg} $\sim 4.3-5.0$.   This
corresponds to a mass of $\sim 10-30$ {\mjup} and an age of $1-5$ Gyr, however there are large
discrepancies between the model and observed spectrum.    The radial and
rotational velocities of the brown dwarf are measured as 46.9~$\pm$~2.5 and
40~$\pm$~10~{\kms}, respectively, reflecting a thin disk Galactic
orbit and fast rotation similar to other T dwarfs, suggesting a young, possibly planetary-mass brown dwarf.
\end{abstract}


\section{Introduction}

Late--type T dwarfs (T $<$ 600 K) are among the dimmest, coldest and
least massive products of star formation.  
As cooling brown dwarfs (BDs), these objects sample a broad range of
age and mass, from old, relatively massive relics of the earliest
epochs of Galactic star formation to recently--formed planetary--mass
objects incapable of deuterium fusion \citep[$\cal{M} <$ 13 {\mjup};][]{2001RvMP...73..719B}. 
  Their numbers in the vicinity of the Sun help constrain the substellar initial mass function
\citep[e.g.,][]{2008ApJ...676.1281M,2010MNRAS.406.1885B}  and the minimum formation
mass \citep[e.g.,][]{2004ApJS..155..191B}.
Late-type T dwarfs also occupy the same physical parameter space
(mass, age, {\teff}) as exoplanets, with moderate
separations (0.5--1 AU) from solar--type stars, making them important
benchmarks of exoplanet models
\citep[i.e.,][]{1997ApJ...491..856B,2003A&A...402..701B} and direct
detection experiments \citep[i.e.,][]{2006SPIE.6272E..18M}.

Despite its astronomical utility, high--resolution spectroscopy of
late T dwarfs is exceedingly rare.  The main culprit is 
faint T dwarf luminosity \citep[$L \sim 2 \times 10^{-6} L_{\odot}$ for T9;][]{2004AJ....127.3516G}.  Recently, surveys
such as the Wide-field
 Infrared Survey Explorer \citep[WISE;][]{2010AJ....140.1868W} and the
 UKIRT Infrared Deep Sky Survey \citep[UKIDSS;][]{2007MNRAS.379.1599L} have produced deep multi--band photometry
 over thousands of square degrees in the
 near-infrared (NIR; $\sim$ 1-2.5$\mu$m) and mid-IR ($\sim$ 3-5
 $\mu$m).  Yet many of the cool T dwarfs discovered in these surveys
 have been studied at low spectral resolutions ($R$ = {\ldl} $< 1000$), blending rich absorption bands
 produced by {\meth}, {\amo} and {\wat} and complicating their
 atmospheric analyses.  These low--resolution observations also limit
 the precision of radial and rotational
 velocity measurements.  Higher resolution observations are only
 feasible for bright objects, biasing their kinematic
 analysis to the nearby BD population.  For example, the largest study of T dwarfs with measured rotation velocities
contained only nine objects \citep{2006ApJ...647.1405Z}.

Currently, there are $< 20$ brown dwarfs classified as T9 or later.
One of the brightest of these sources is
{\name}  (hereafter {\namesh}), discovered in the UKIDSS data by
\cite{2010MNRAS.408L..56L}, and tentatively assigned a spectral type of T10 based
on its strong molecular absorption and faint absolute magnitude.   The spectral type of {\namesh}
was revised by \cite{kirkpatrick} and \cite{cushing} to T9 and it has
been designated as the infrared spectral standard.  {\namesh} is a tantalizing target for
followup studies, as it is relatively nearby ($d \sim$ 4.1 pc; see
Table \ref{table:properties}), bright
($J = 16.5$).   
In this paper, we present a moderate resolution (R $\sim$ 6000) NIR
spectrum of {\namesh}, acquired using the newly installed Folded-Port Infrared Echellette Spectrograph
\citep[FIRE;][]{2010SPIE.7735E..38S}.  The observations are detailed in
Section \ref{sec:obs}.  In Section \ref{sec:results}, the NIR SED and corresponding model fits are
shown along with the rotational and radial
velocities and Galactic orbit of {\namesh}. Finally, our
conclusions and paths for future investigations are presented in Section
\ref{sec:conclusions}.

\section{FIRE Observations and Reductions}\label{sec:obs}

The FIRE spectrograph \citep[][]{2008SPIE.7014E..27S,2010SPIE.7735E..38S} was installed and commissioned on the Baade Magellan
telescope at Las Campanas Observatory during March and April 2010.  FIRE is a
single-object spectrograph with two modes: a cross--dispersed echellette
mode with moderate resolution ($R \sim 6000$), and a longslit low
resolution mode ($R \sim 250-350$).   The spectrum is imaged on a
HAWAII-2RG chip, with continuous coverage from 0.85 - 2.5 $\mu$m.  In
the cross--dispersed mode, the spectrum is spread over 21 orders, with
some overlap in wavelength coverage at the edges of each order.
Target acquisition is achieved with a second NIR imager and Mauna Kea Observatory $J$ filter
focused on the entrance slit.  
FIRE was designed to be sensitive,
employing the latest generation of HgCdTe detectors while limiting the
number of reflective and transmissive surfaces, resulting in a
zero-point of $\sim 16$ mag (for 1~count~pixel$^{-1}$~s$^{-1}$ across $JHK$) for
the echelle mode.  

On April 6, 2010 UT, we obtained four 900s exposures of
{\namesh} in FIRE's echelle mode.  The sky was clear
with no cloud cover, and seeing was
$\sim 0^{\prime\prime}.5$ in $J$ at the time of observation.  
The 0$^{\prime\prime}.6$ slit was used and
aligned with the parallactic angle and the airmass was $1.4$.  The exposures
were dithered along the slit in a ABBA pattern and a Fowler sampling
of 8 was employed.  An A0V star, HIP 63714, was observed for telluric correction and flux calibration purposes.  Quartz flat fields and thorium-argon (ThAr) arcs were obtained after
the science and telluric calibrator exposures.   

\begin{figure*}[htbp]
\centering
\includegraphics[scale=0.7]{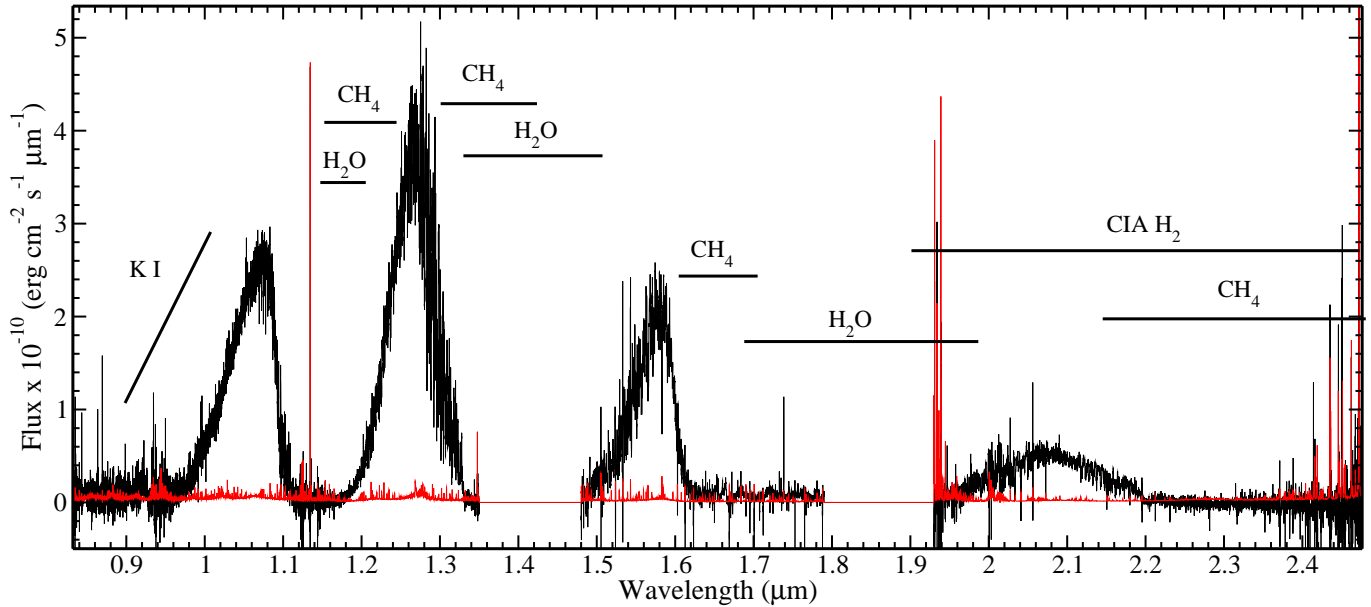} 

 \caption{FIRE spectrum of {\namesh} (black line).  The noise spectrum is shown in
   red.  Major atomic and molecular
   absorption features are labeled.  Note the strong {\meth} absorption,
   indicative of a cool T dwarf.  Strong telluric absorption between bands have been masked.}
   \label{fig:spectrum}
\end{figure*}

The images were reduced using the FIRE reduction software package,
FIREHOSE, which is based on the MASE pipeline \citep{mase}
for the MagE spectrograph \citep{2008SPIE.7014E.169M}.  FIREHOSE, like
MASE, was designed to reduce cross--dispersed echelle spectra with
curved orders.   Quartz lamp images were used to identify the order
boundaries and derive flat-field and illumination corrections.
A combination of OH telluric
lines and ThAr arc images were used to determine the wavelength
solution along the center of each order and its tilt in the spatial
direction, which was used to
construct a two dimensional vacuum wavelength map.  The typical uncertainty
of the wavelength solution was 0.15 pixels, corresponding to 0.04-0.4 Angstroms
depending on the order.  A 2D sky model was constructed using basis splines
\citep{2003PASP..115..688K} and subtracted from each order.  This step eliminates
the need for ABBA-type dithers for sky subtraction\footnote{However,
  for faint sources, multiple exposures are still needed to avoid saturating sky
  lines.}. An
optimal extraction routine was then performed on each
order \citep{1986PASP...98..609H}, extracting the object flux onto a
heliocentric rest frame wavelength grid.  Telluric corrections were incorporated into the
pipeline using a modified version of \textit{xtellcor} from Spextool
\citep{2004PASP..116..362C, 2003PASP..115..389V}.  Telluric absorption was quantified by
comparing the telluric stellar spectrum to a model Vega spectrum reddened using the $B-V$ color of HIP 63714, velocity shifted, and broadened using the 1.005~$\micron$ H~I Pa~$\delta$ line as a line kernel.  The
science target was also flux calibrated during this step.  Multiple
spectra of the same target were combined after flux calibration.  Finally, the
extracted orders were combined into a 1d spectrum, with overlap regions
averaged together.  The
final spectrum is shown in Figure \ref{fig:spectrum}.  The
peak signals to noise in the $y, J, H$ and $K$ bands are $\sim$
250, 350, 200, and 60 respectively.  As a test of the flux calibration
and telluric correction, we computed synthetic $J-H$ and $H-K$ colors
from the final spectrum, and compared them to the measured values
reported in Table \ref{table:properties}.  The $J-H$ color agreed
within 0.1 mag, while $H-K$ differed by $\sim$ 0.5 mag, indicating the
overall flux calibration between orders is sufficient for spectral analysis.

\begin{center}
\begin{deluxetable}{lrl}
\tablewidth{0pt}
 \tablecaption{Measured Properties of {\namesh}}
 \tabletypesize{\scriptsize}

 \tablehead{
 \colhead{Property} &
 \colhead{Value} &
 \colhead{Source\tablenotemark{a}}
}
 \startdata
$\alpha$ (J2000)& 07:22:27.51 &1 \\
$\delta$ (J2000) & -05:40:31.2 & 1\\
$\mu_{\alpha}$ (mas yr$^{-1}$)  & -910 $\pm$ 8 & 1 \\
$\mu_{\delta}$ (mas yr$^{-1}$) & 1020 $\pm$ 3 & 1 \\
$\pi$ (mas) & 237 $\pm$ 41  & 1 \\ 
RV ({\kms}) & 46.9 $\pm$ 2.5 & 2 \\
$v_{\rm tan}$ ({\kms}) & 19 $\pm$ 4 & 1 \\
{\vsini} ({\kms}) & 40 $\pm$ 10  & 2 \\
$V_R$ ({\kms}) & -42 $\pm$ 2 & 2 \tablenotemark{b}\\
$V_{\phi}$ ({\kms}) & 221 $\pm$ 1 & 2 \tablenotemark{b} \\
$V_Z$ ({\kms}) & 4 $\pm$ 1 & 2  \tablenotemark{b} \\
$i$   & 24.80 $\pm$ 0.13   & 1\\
$z$   & 20.51 $\pm$ 0.09   & 1\tablenotemark{c}\\
$Y$   & 17.37 $\pm$ 0.02   & 1\\
$J$   & 16.52 $\pm$ 0.02   & 1\\
$H$   & 16.90 $\pm$ 0.02 & 1\\
$K$   & 17.07 $\pm$ 0.08 & 1\\
$[3.6]$   & 14.28 $\pm$ 0.05    & 1\\
$[4.5]$   & 12.19 $\pm$ 0.04   & 1\\
$W1$   & 15.15 $\pm$ 0.05    & 4\\
$W2$   & 12.17 $\pm$ 0.03    & 4\\
$W3$   & 10.18 $\pm$ 0.06    & 4\\
Spectral Type & T9 & 3 \\
{\teff}   & $500-600$ K    & 2\\
{\logg}   & $4.2-5.0$    & 2\\
Mass  & $10-30$ {\mjup}    & 2\\
Age  & $1-5$ Gyr    & 2\\
\enddata

 \label{table:properties}
\tablenotetext{a}{1~-~\cite{2010MNRAS.408L..56L}, 2~-~This Paper, 3~-~\cite{cushing}, 4~-~\cite{2010AJ....140.1868W}}
\tablenotetext{b}{Computed using the solar velocity of \cite{2010MNRAS.403.1829S}.}
\tablenotetext{c}{Average of two $z$ reported values.}

\end{deluxetable}
\end{center}

\section{Results}\label{sec:results}
\subsection{Spectral Properties}

\begin{figure*}[htbp]
$
\begin{array}{c}
\includegraphics[scale=0.4]{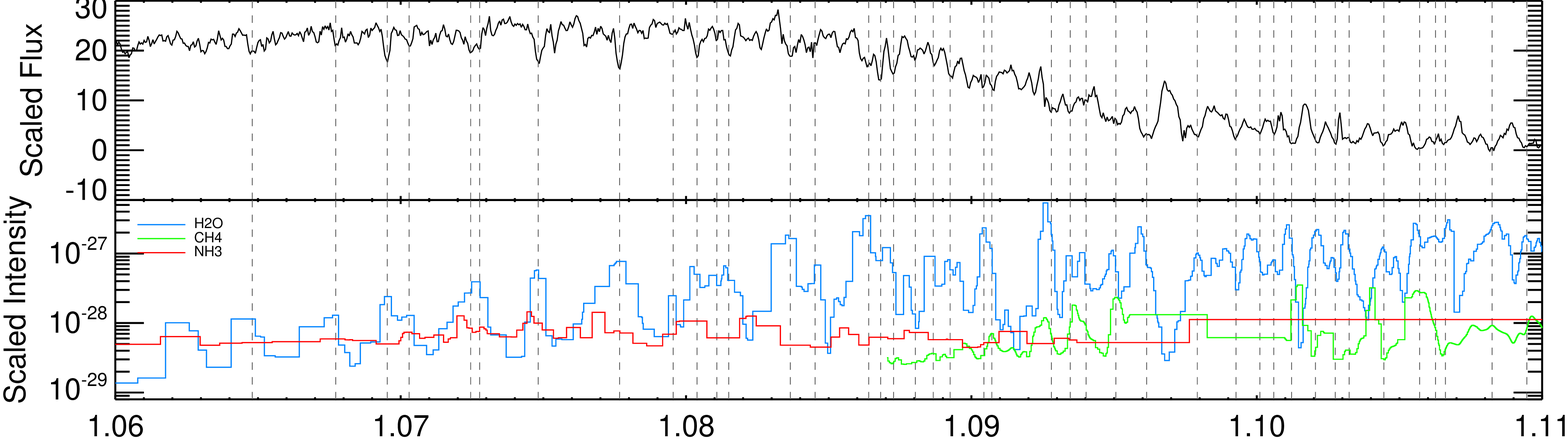} \\
\includegraphics[scale=0.4]{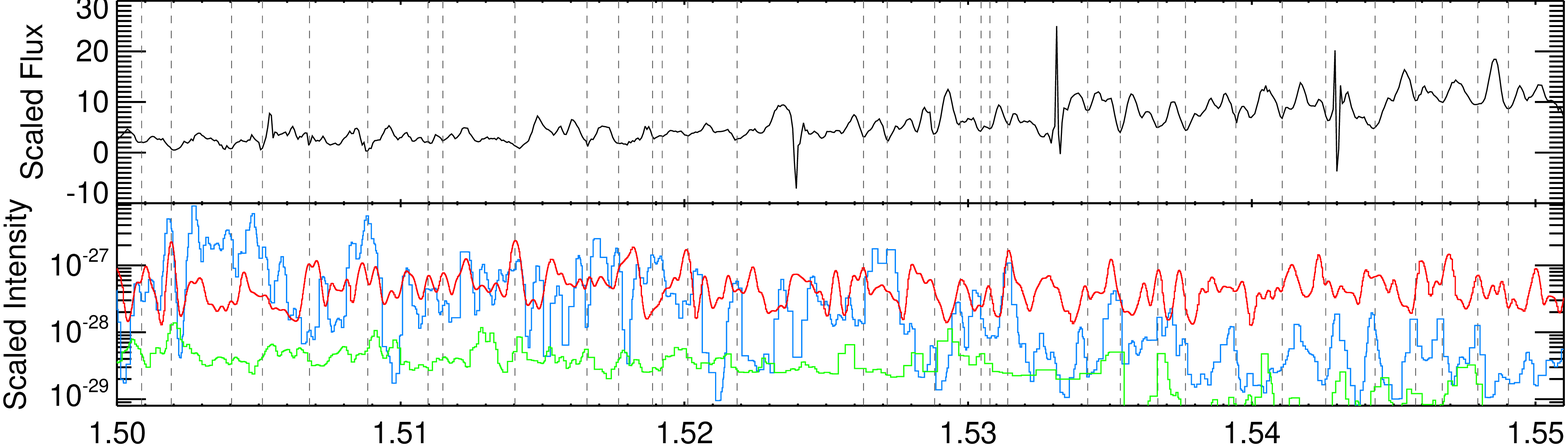} \\
\includegraphics[scale=0.4]{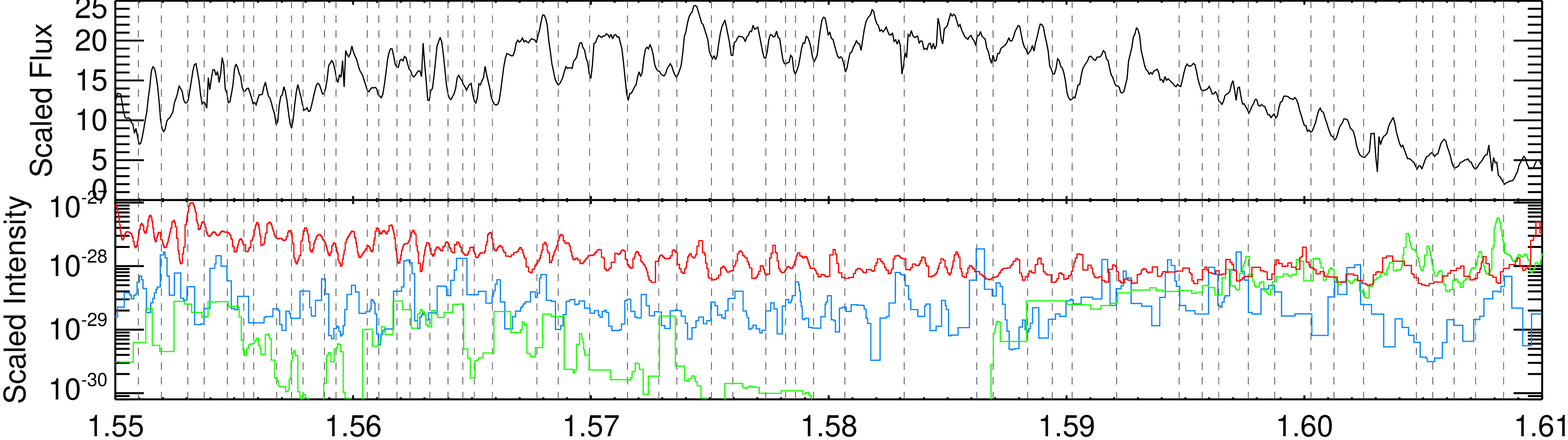} \\
\includegraphics[scale=0.4]{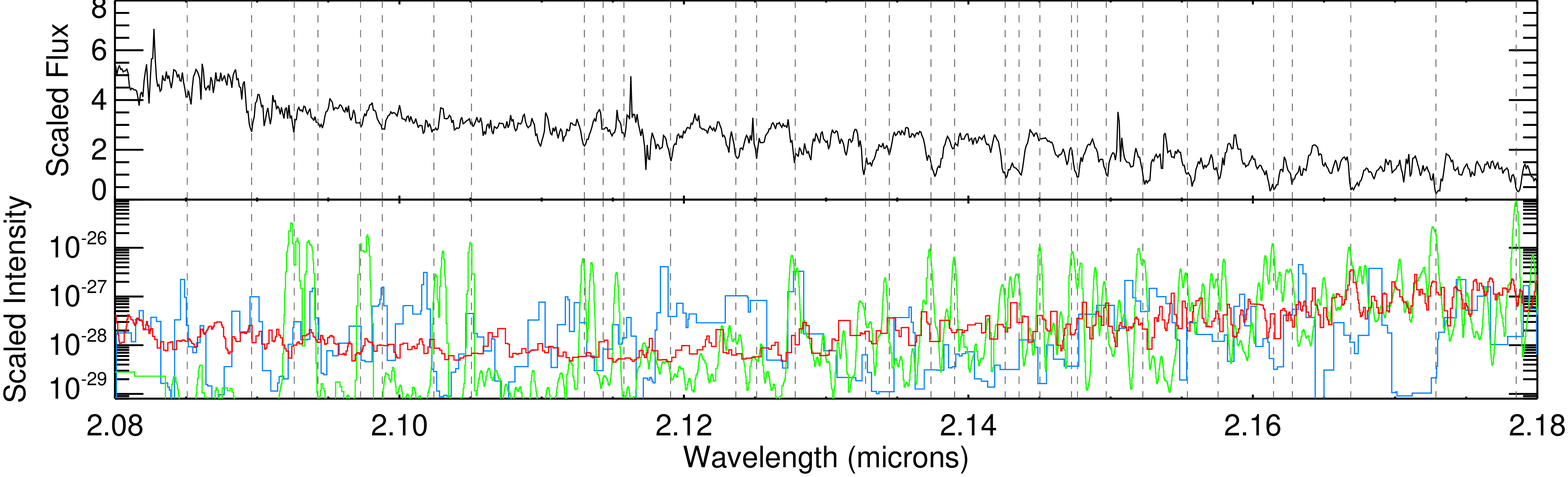} 
\end{array}$
 \caption{Comparison of {\namesh}'s spectrum (black line) to molecular
   linelists ({\amo}: red line, {\meth}: green line, {\wat}: blue
   line).  The absorption intensity (cm molecule$^{-1}$)  was scaled
   by the relative abundances for each molecule at $\log~T$ = 2.7 
   \citep[from][]{2006ApJ...647..552S}.  Absorption features detected
   in both the spectrum of 
   {\namesh} and the molecular absorption intensity spectra are marked by vertical
   dashed lines.
   }
   \label{fig:linelist}
\end{figure*}

Since the FIRE spectrum of {\namesh} is the highest resolution
observation of one of the coolest BDs, we compiled the most
recent laboratory line lists for {\wat}, {\meth} and {\amo} to
identify absorption features
\citep[][respectively]{2006MNRAS.368.1087B,2003JQSRT..82..279N,2011MNRAS.413.1828Y}.
The HITRAN 2008 database was also used to supplement the linelists \citep{2009JQSRT.110..533R}.
Prior to comparing to the FIRE spectrum of {\namesh}, each linelist
was cropped to 0.8-2.5 $\mu$m and smoothed
with a 50 {\kms} Gaussian kernel, which corresponds to one FIRE
resolution element.  The absorption intensities for each linelist (in
units of cm molecule$^{-1}$) were scaled by the relative molecular abundances
shown in Figure 3 of \cite{2006ApJ...647..552S} for {\teff} = 500 K.
The non--equilibrium abundances were assumed for {\amo}.
 The scaled values were then plotted along with the {\namesh} spectrum, and molecular features
were manually identified.  An example of our linelist comparisons are
shown in Figure \ref{fig:linelist}.

\begin{figure*}[htbp]
\centering
\includegraphics[scale=0.7]{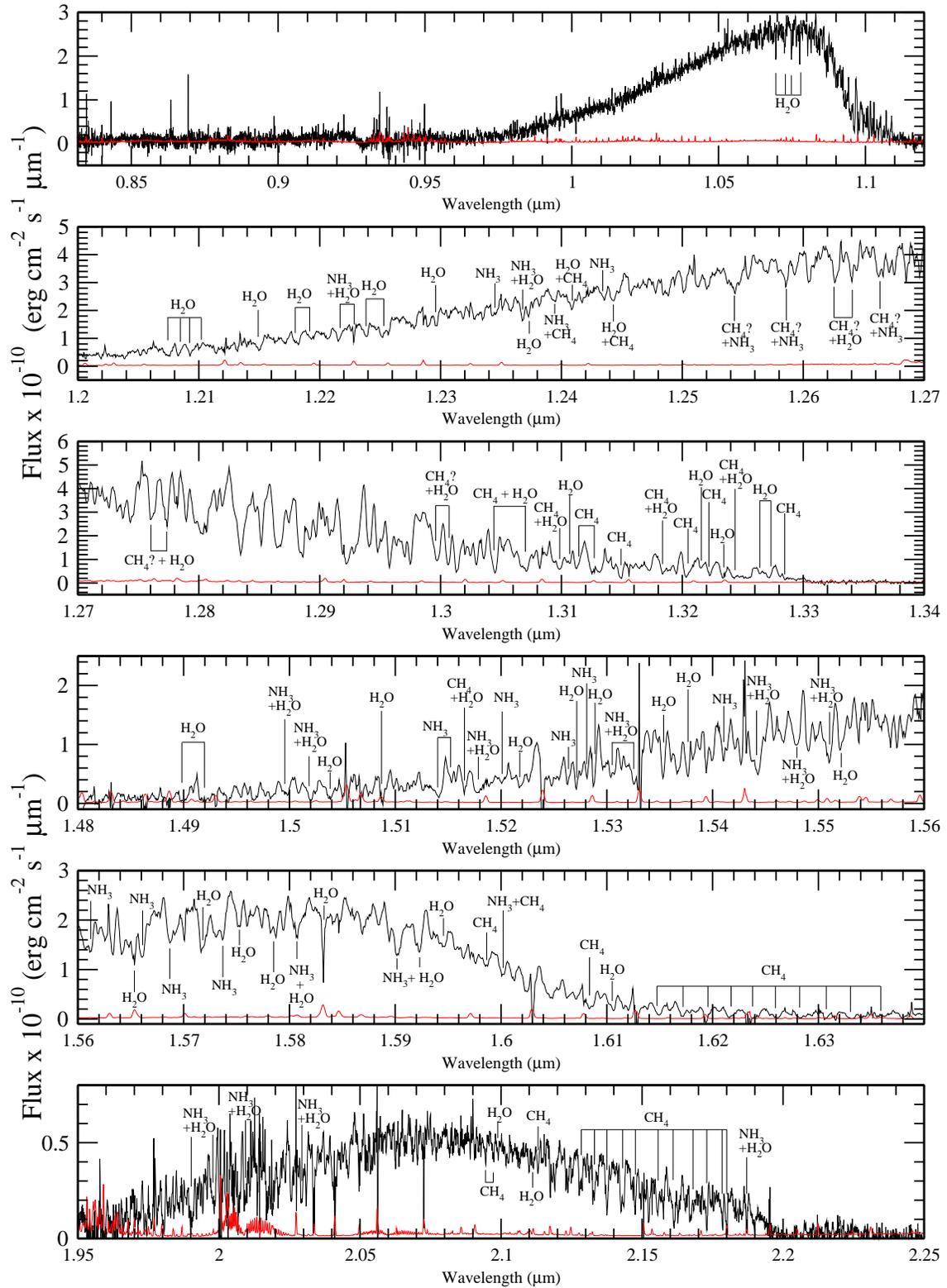} \\
\caption{FIRE spectrum of {\namesh} (black line). The noise spectrum is shown in
   red.   Major molecular features are labeled.  Some {\meth} features
 are labeled with a ``?'' indicating regions where linelists are
 incomplete, but strong methane absorption has been observed \citep{1979ApJ...233.1021F}.}
   \label{fig:spectrum_zoom}
\end{figure*}

In Figure \ref{fig:spectrum_zoom}, we plot expanded views of the
{\namesh} spectrum in the $y, J, H,$ and $K$ bands.  Prominent absorption
features are labeled in each panel.   Of note in Figure \ref{fig:spectrum_zoom} is the significant number of
absorption features throughout the spectrum that are attributed to
{\wat}, {\meth} and {\amo}.  There are a large number of blends
between the molecular features, especially with {\wat}, but some
isolated absorption bands do exist. We confirm the tentative
identification of {\amo} by \cite{2010MNRAS.408L..56L} near 1.514
$\mu$m.  Additional isolated {\amo} absorption features can be found
near 1.234, 1.244, 1.52, 1.526, 1.542, 1.56, 1.566, 1.568 and 1.574
$\mu$m.   This suggests that observations spanning 1.5 - 1.6 $\mu$m
present the best chance at directly detecting {\amo} at these
temperatures.  While the detection of {\amo} has been suggested as the
hallmark of the Y spectral class \citep[i.e.,][]{2007ApJ...667..537L},
these weak features are consistent with the end of the T dwarf sequence as
advocated by \cite{cushing}.

{\meth} exhibits prominent
absorption bands near 1.6 and 2.15 $\mu$m.   The structure seen in the
{\namesh} spectrum is usually not detected at lower resolutions.
These features may be used to derive spectral indices and provide
isolated regions of the spectrum to derive atmospheric parameters (see
Section \ref{sec:parameters}).

\subsection{Radial Velocity}\label{sec:rv}
The resolution of FIRE's echelle mode permits
the measurement of {\namesh}'s radial velocity (RV) with a precision
of a few {\kms}.  We
cross--correlated the spectrum of {\namesh} against other T dwarfs
observed with FIRE and model atmosphere predictions.  The T dwarf RV
standards (Table \ref{table:rv})
were taken from \cite{2007ApJ...666.1205Z} and were observed with a
similar setup on FIRE as
part of a larger effort to quantify BD kinematics (Burgasser et al., in preparation).   We
also employed the T= 400 K, 500 K, 600 K and 700 K models (with
{\logg} = 5.0, $[m/H] = 0.0$) from the BT-Settl
grid \citep{2010arXiv1011.5405A}.   The 1.27-1.31 $\mu$m spectral
region in the $J$ band contains many strong molecular features, making it ideal for
cross--correlation.  The models were smoothed to match
the resolution of the FIRE observations.  Cross--correlations were
computed using the \textit{xcorl} IDL routine
\citep{1999AJ....118.2460B, 2003ApJ...583..451M, 2009ApJ...693.1283W}.  The reported RV and uncertainty were
computed by taking the unweighted
mean and the standard deviation of the individual RV measurements, respectively.  The
measured RV and uncertainty for {\namesh} is 46.9 $\pm$ 2.5 {\kms}.  

\subsubsection{Galactic Orbit}
To frame the kinematics of {\namesh} in a Galactic context, we
computed its orbit using its measured velocity and position as initial
conditions.  Using the distance, position, proper motion and radial
velocity reported in Table \ref{table:properties}, we computed the
cylindrical velocity vector [$V_R$, $V_{\phi}$, $V_Z$] where the local
standard of rest is [0, 220, 0] {\kms} \citep{1986MNRAS.221.1023K},
the solar motion is [11.1, 12.24, 7.25] {\kms} \citep{2010MNRAS.403.1829S} and the radial velocity
component $V_R$ increases in the direction of the Galactic center \citep{1987AJ.....93..864J}.
The Sun's radial position ($X$) was taken to be 8.5 kpc away from the Galactic
center \citep{1986MNRAS.221.1023K} and 27 pc above the Plane
\citep[Z;][]{2001ApJ...553..184C,2008ApJ...673..864J}.  Note that we are reporting 
velocities in a Galactocentric frame, rather than the traditional
heliocentric $UVW$ frame.  While the effect is negligible for {\namesh}, using $UVW$ rather than a Galactic frame can introduce errors
of a few {\kms} for distances $\gtrsim$ 100 pc.   Employing a
Galactocentric velocity frame will be important as more distant dwarfs
are discovered in the next generation of surveys (i.e., LSST).

The orbit was integrated assuming a set of static,
spherically-symmetric oblate Plummer's sphere potentials for the
Galactic halo, bulge and disk, using the forms described
in \cite{kuzmin1956} and \cite{1975PASJ...27..533M} and with parameters
from \cite{1995A&A...300..117D}.    A Runge-Kutta integrator was used to calculate the
orbit over a period of $\pm$250~Myr with a 10~kyr timestep, and both
energy and the $Z$-component of angular momentum were conserved to
better than one part in 10$^{3}$.  To sample measurement
uncertainties in the distance and velocity of {\namesh} relative
to the Sun, we computed 100 realizations of the orbit through the
Monte Carlo method, varying the starting conditions assuming normal
distributions with means and widths given by the values in Table \ref{table:properties}.

\begin{figure*}[htbp]
\centering
\includegraphics[scale=0.4]{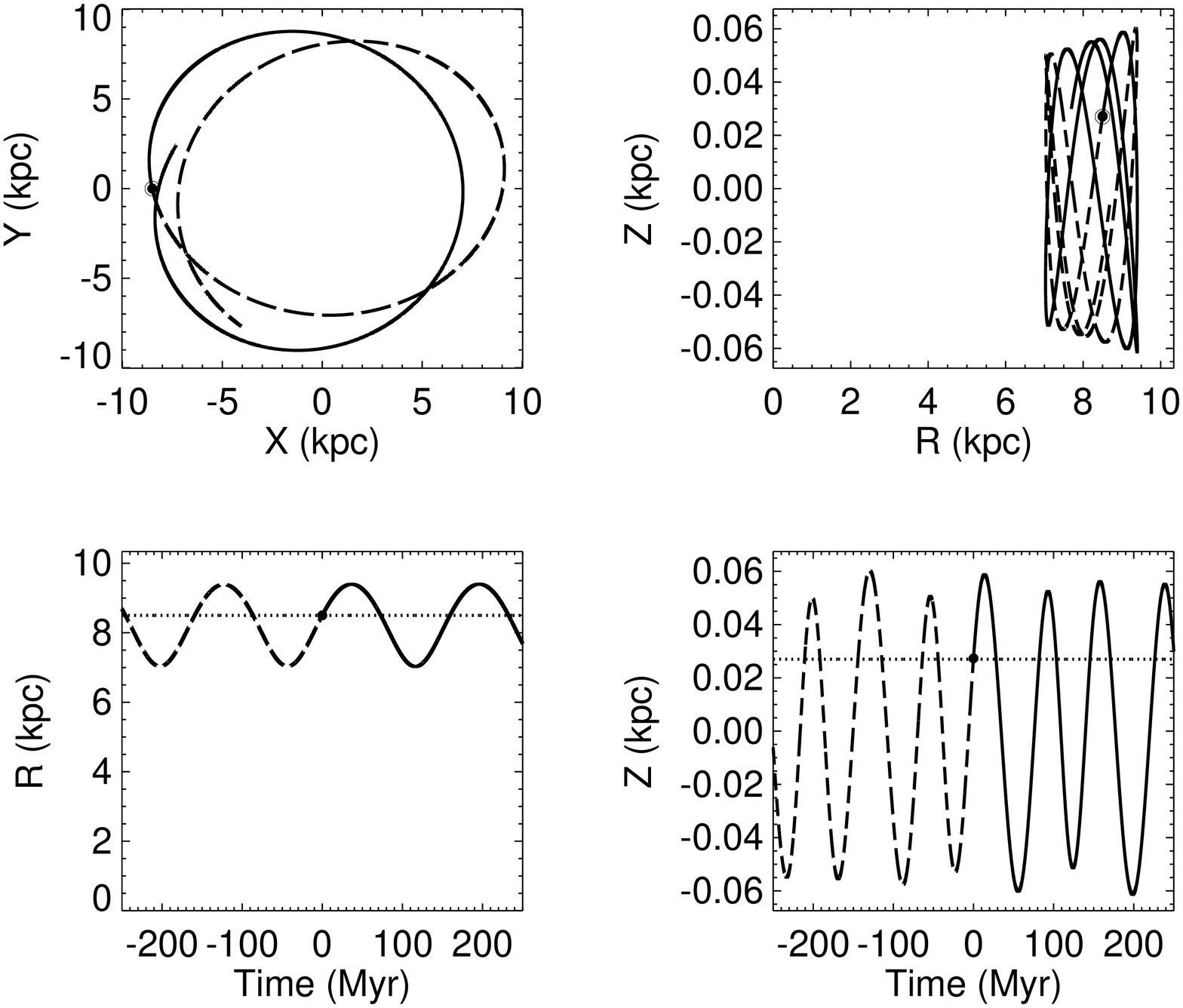} 

 \caption{Simulated orbit of {\namesh}.  The object maintains a low eccentricity
 thin disk orbit over the length of the simulation.  The radial
 excursions of {\namesh} are $\sim$ 1 kpc, while the object maintains a vertical
 displacement of $\lesssim$ 60 pc away from the Galactic plane.}
   \label{fig:orbit}
\end{figure*}

The baseline calculation is shown in Figure \ref{fig:orbit}, revealing a flat orbit
with small eccentricity ($e$).  The maximum vertical displacement of the source
from the Galactic plane never exceeds ${Z}~\sim~60$~pc, with 
radial excursions between 7 kpc $< R <$ 9.5 kpc, and $e$ = 0.11$\pm$0.02.  This orbit is consistent with membership
in the Galactic thin disk population \citep{2007AJ....134.2418B} suggesting that {\namesh} is a relatively young brown dwarf.
However, we strongly caution the use of
kinematic properties as an age discriminant, as they should only be
considered in a statistical manner.  
Computing orbits for larger samples of MLTY
dwarfs will help place the orbit of {\namesh} in a broader context.

\subsection{Rotational Velocity}
\begin{figure}[htbp]
\centering
\includegraphics[scale=0.60]{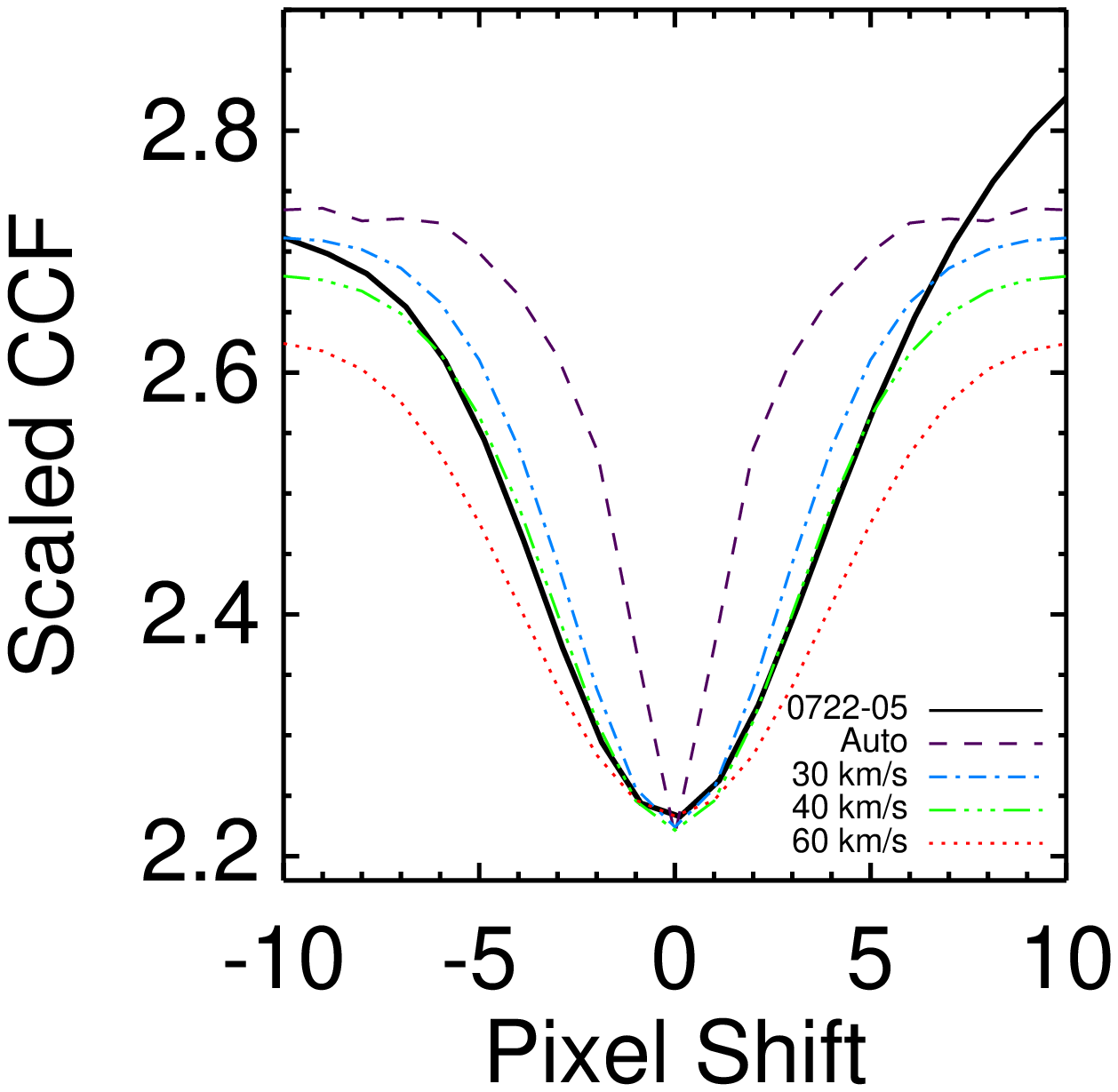} 

 \caption{The cross-correlation functions of the broadened T = 500 K
   models.  The purple dashed line corresponds to the auto--correlation
   function, the blue dash-dot line is the {\vsini} = 30 {\kms} template, the
   green dot-dot-dash line is the {\vsini} = 40 {\kms} template and
   the red dotted line is the {\vsini} = 60 {\kms} template.  The CCF
   of {\namesh} with the unbroadened template is the solid black
   line.  The CCFs have been shifted and scaled to match at the minimum.
  It is evident that there is some noticeable rotation, as the
  auto-correlation function is narrow compared to the observation.
  In contrast, the {\vsini} = 60 {\kms} is too wide.  We report the
  {\vsini} of {\namesh} as 40 $\pm$ 10 {\kms}.}
   \label{fig:vsini}
\end{figure}

The rotational velocity of  {\namesh} was also measured with a
technique used by a number of previous studies
\citep[e.g.,][]{2000MNRAS.316..827R, 2003ApJ...583..451M,
  2009ApJ...693.1283W}.   Briefly, the
science object ({\namesh}) is cross--correlated with a rotationally unbroadened template
(the T = 500 K, {\logg} = 5.0 model, $[m/H] = 0.0$ from BT-Settl,
\citealp{2010arXiv1011.5405A}).  The model is convolved with
a Gaussian of 50 {\kms} to approximate the effects of FIRE's line spread function.
This cross--correlation function (CCF) is compared to the CCFs derived
from correlating the unbroadened template to rotationally broadened versions of
itself.  We constructed a series of rotating templates ranging in
{\vsini} from 10 to 120 {\kms}, in steps of 10 {\kms} using the technique described in \cite{1992oasp.book.....G}.  In Figure \ref{fig:vsini}, we compare the
auto--correlation of the template against the CCFs of the {\vsini} = 30,
40 and 60 {\kms} templates and the CCF of {\namesh} over the same
wavelength range described in Section \ref{sec:rv}.  
While {\vsini}  = 40 {\kms} was the closest match to the CCF of
{\namesh}, the {\vsini} = 30 and 50 {\kms} were good fits to most of
the CCF trough (see Figure \ref{fig:vsini}).  Thus, we report a {\vsini} for
{\namesh} is $40 \pm 10$ {\kms}.  

This rotation velocity is similar to
those of late-L and T dwarfs
\citep[e.g.,][]{2006ApJ...647.1405Z,2008ApJ...684.1390R}.  Of the nine
T dwarfs observed by \cite{2006ApJ...647.1405Z}, only one
brown dwarf (SDSS$~$J134646.45$~-$003150.4), had a {\vsini} under 20
{\kms} and the authors speculated that this may be due to
inclination\footnote{Using the binomial function formalism from
  \cite{2010AJ....139..504B}, there is a $\sim 72\%$ chance of observing one
  slow rotator ($< 20$ {\kms}) in a sample of nine stars, assuming they
  all rotate at 40 {\kms}.}.
The remaining 8 T dwarfs demonstrated rotational velocities between 20 and 40
{\kms} and this distribution does not vary significantly from the
observed rotation velocities of L dwarfs \citep{2008ApJ...684.1390R}.  The {\vsini} of {\namesh} reinforces the findings of
previous studies suggesting that T dwarfs are inefficient
at rotational braking.  In solar--type stars, magnetic fields power
two sources of angular momentum loss:  disk braking and flaring
events \citep{1972ApJ...171..565S,1997A&A...326.1023B}.  Large--scale magnetic fields can form in
convective, rotating low--mass objects
\citep[i.e.,][]{2008ApJ...676.1262B} and have been observed in some
late M and  L
dwarfs \citep{2008ApJ...684.1390R, 2008ApJ...684..644H}, but remain
undetected in T dwarfs \citep{2006ApJ...648..629B}.    This may
indicate the lack of magnetic fields in T dwarfs, or alternatively, a weak
coupling between the fields and the predominately neutral atmosphere.
This de-coupling would reduce disk braking and flare frequency (due to
reconnection events), which would decrease angular momentum loss in T
dwarfs compared to solar--type stars.

\begin{center}
\begin{deluxetable*}{lll}
\tablewidth{0pt}
 \tablecaption{Radial Velocity Measurements}
 \tabletypesize{\scriptsize}

 \tablehead{
 \colhead{Standard} &
 \colhead{RV ({\kms})} &
 \colhead{Notes\tablenotemark{a}}
}
 \startdata
T = 400 K  & 47.9 & BT-Settl model from \cite{2010arXiv1011.5405A} \\
T = 500 K  & 49.2 & BT-Settl model from \cite{2010arXiv1011.5405A}  \\
T = 600 K  & 48.5 & BT-Settl model from \cite{2010arXiv1011.5405A}  \\
T = 700 K  & 48.5 & BT-Settl model from \cite{2010arXiv1011.5405A}  \\
2M0415-0935 & 45.6 &  Assuming 49.6 {\kms} for template \citep{2007ApJ...666.1205Z} \\
2M0559-1404 & 46.8 &  Assuming -13.8 {\kms} for template \citep{2007ApJ...666.1205Z} \\
2M1553+1532 & 41.8 & Assuming -32.9 {\kms} for template \citep{2007ApJ...666.1205Z} \\
\hline
Mean RV & 46.9 & 2.5 {\kms} uncertainty\\
\enddata

 \label{table:rv}
\tablenotetext{a}{All RV measurements were computed in the $J$ band
  from 1.27 to 1.31 $\mu$m.}
\end{deluxetable*}
\end{center}

\subsection{Atmospheric Model Fits}\label{sec:parameters}

To examine the physical properties of {\namesh}, we compared its FIRE
spectrum to the BT-Settl models of \citet{2010arXiv1011.5405A}. These
models are based on the PHOENIX code \citep{1999ApJ...525..871H}, and
reflect an update to the original Settl models of
\citet{2003IAUS..211..325A} with a microturbulence velocity field
determined from 2D hydrodynamic models \citep{2010A&A...513A..19F} and
updated solar abundances from \citet{2009ARA&A..47..481A}.  We adopted
a set of solar-metallicity ([$m/H$] = 0.0) models sampling {\teff} = 400--900~K in
100~K steps, and {\logg} = 3.5--5.5~{\cmss} in 0.5~{\cmss} steps, with
the exceptions that violate evolutionary parameters (e.g., {\teff}
$\le$ 700~K and {\logg} = 5.5).

Our fitting procedure was based on the formalism developed by
\cite{2008ApJ...678.1372C}, \cite{2009ApJ...706.1114B} and
\cite{2010ApJ...725.1405B}.  Model surface fluxes (in $f_{\lambda}$
units) were smoothed to a common resolution of {\ldl} = 6000 using a
Gaussian kernel, and both models and FIRE data were interpolated onto
a common wavelength grid spanning 0.9 to 2.4~$\micron$.  The FIRE data
were also scaled to the observed $J$ magnitude of
{\namesh}.  We then performed eight separate fits to the data,
encompassing the full spectral range (excluding regions of strong
telluric absorption), the $yJHK$ spectral peaks, and three ``narrow''
regions (0.04--0.12~$\micron$ in width) sampling strong molecular
absorption (Table~\ref{table:model}).  Data and models were compared using a
$\chi^2$ statistic, with the degrees of freedom equal to the number of
resolution elements sampled.  The optimal scaling factor minimizing
$\chi^2$ was computed following \citet{2008ApJ...678.1372C}, and is
equivalent to $(R/d)^2$ where $R$ is the radius of the brown dwarf and
$d$ its distance from the Sun \citep{2009ApJ...706.1114B}.  Two sets
of fits were done, one in which the distance was treated as a free
parameter and one in which the model-derived distance must agree with
the parallax measurement of \citet{2010MNRAS.408L..56L} to within
5$\sigma$.  We also allowed for variations in the radial
($\pm$50~{\kms} in steps of 1.25 {\kms} about the reported value) and rotational velocities
(0--100~{\kms} in steps of 3 {\kms}) of the model templates to find a
$\chi^2$ minimum.  Means and uncertainties in the atmospheric parameters ({\teff},
{\logg}) and associated physical parameters (mass, age,
and radius based on the evolutionary models of
\citealt{2003A&A...402..701B}) were determined using the F-test
probability distribution function (F-PDF) as a weighting factor, as
described in \citet{2010ApJ...725.1405B}.  We also propagated sampling
uncertainties of 50~K and 0.25~dex for {\teff} and {\logg},
respectively.

\begin{figure*}[htbp]
\centering$
\begin{array}{cc}
\includegraphics[scale=0.3]{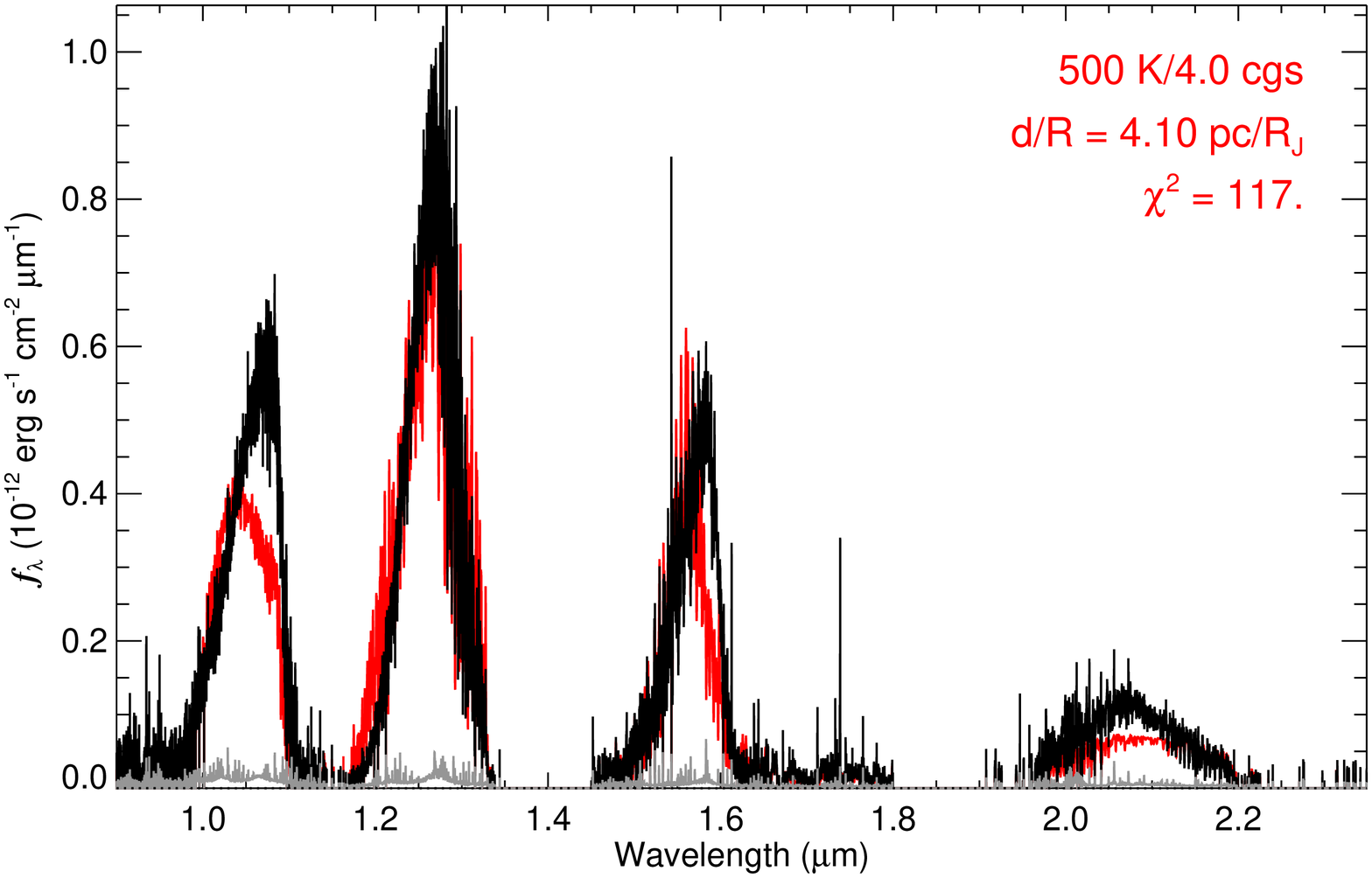} &
\includegraphics[scale=0.3]{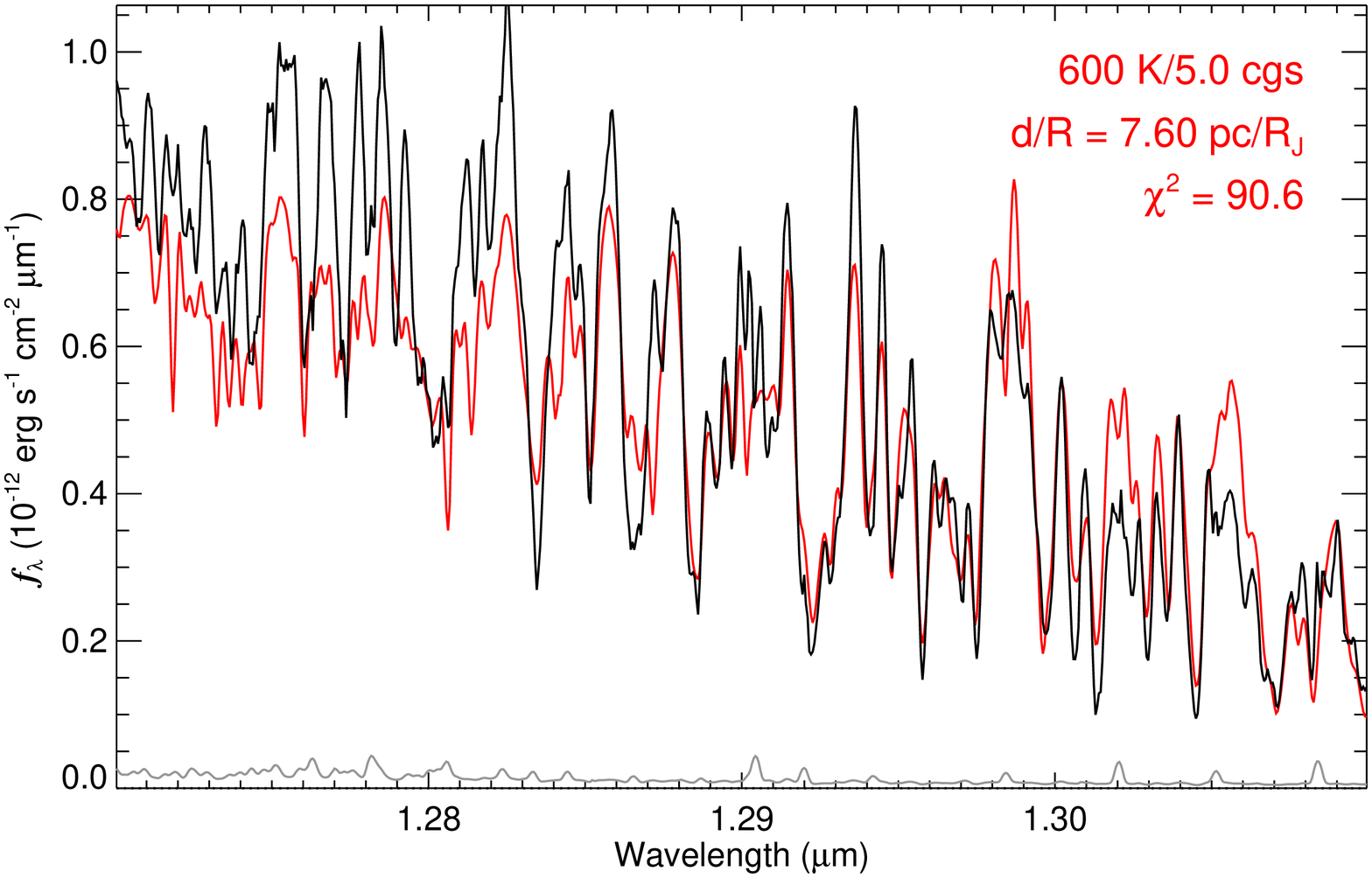}\\ 
\includegraphics[scale=0.3]{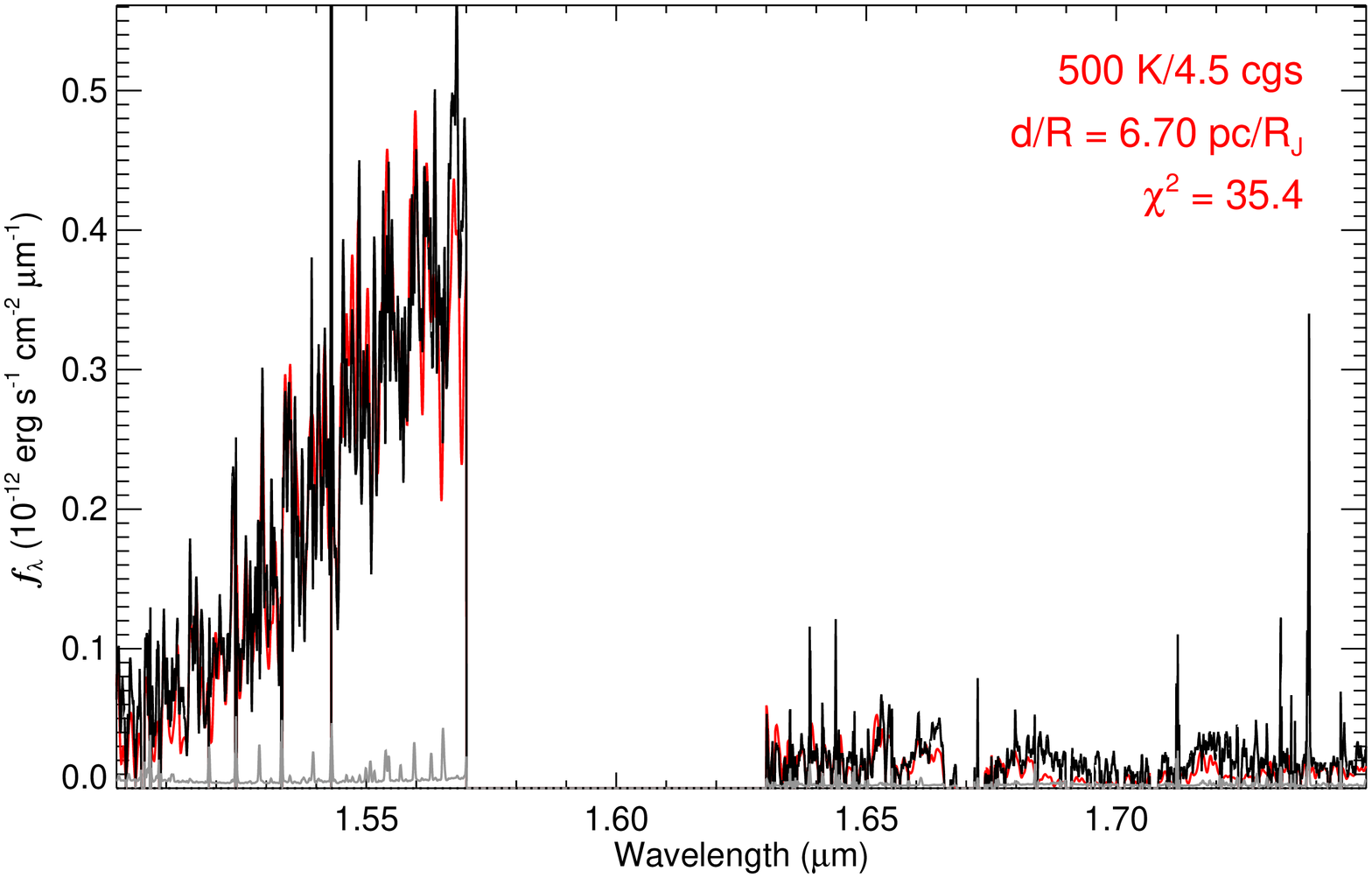} &
\includegraphics[scale=0.3]{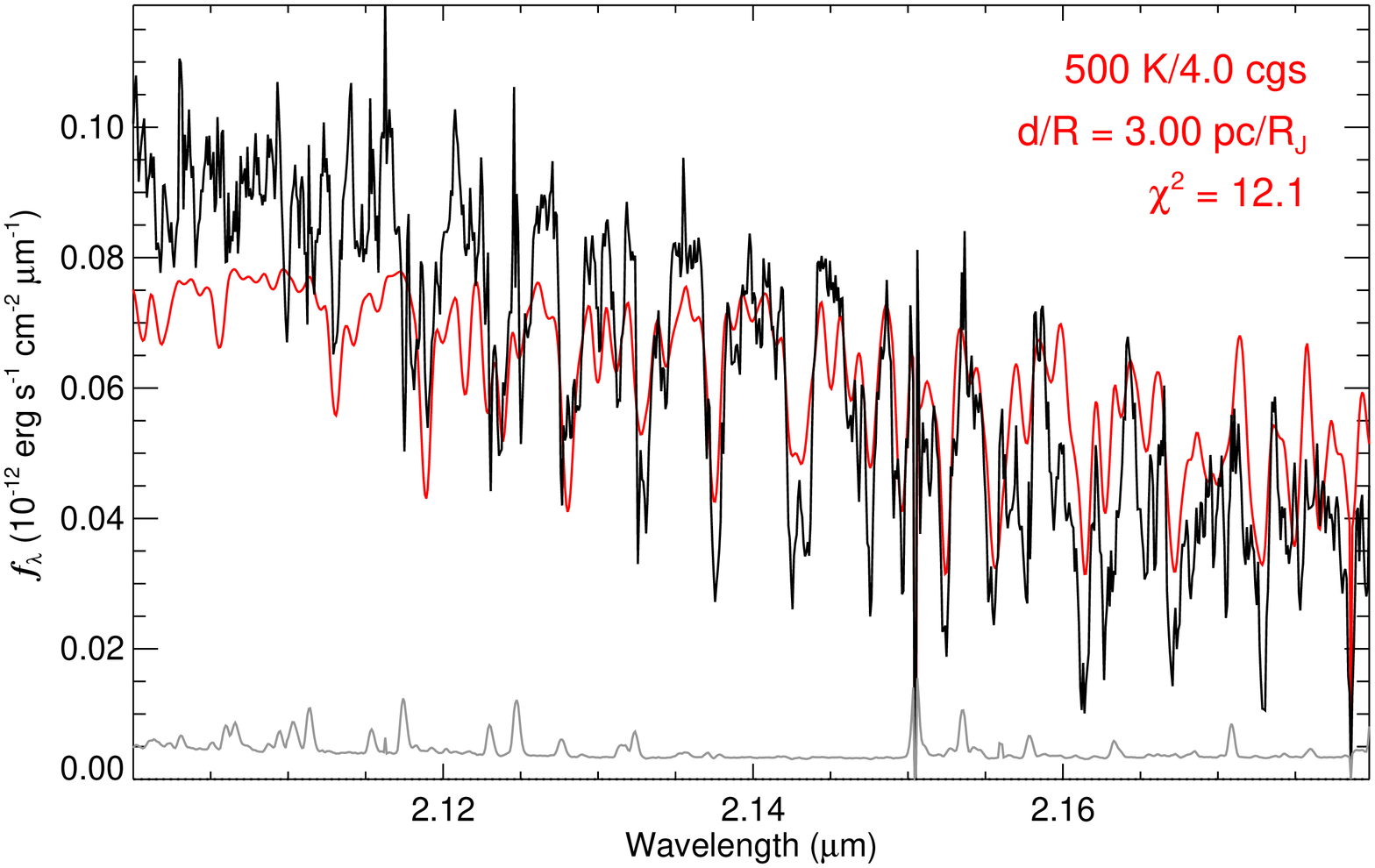} 
\end{array}$
 \caption{Shown are the best fit model atmospheres (red) compared to
   the NIR spectrum of {\namesh} (black line) and error spectrum (gray
   line) for the entire spectrum (upper left panel), and narrow
   regions in the $J$ band (upper
   right panel), $H$ band (lower left panel) and $K$ band (lower right
   panel) for the distance restricted fits.  The agreement between the best fit model (T = 500 K,
   {\logg} = 4.0, $[m/H]$ = 0.0) and the FIRE observations (upper left
   panel) is best near the $J$ band
   peak, but there are large deviations throughout the NIR regime.
   However, within restricted wavelength ranges (other panels) the
   agreement between the model and data improves. }
   \label{fig:model_fit}
\end{figure*}

The best--fit models and data are plotted in Figure
\ref{fig:model_fit}.  The upper panel displays the best fit to the
entire spectrum, with the distance limit enforced.  There are
significant deviations between the model and data in the $JHK$ bands, suggesting that
there remain missing or incorrect molecular opacity in the BT-Settl model calculations.
However, the agreement between the best--fit distance--restricted model and data improves
in the ``narrow'' regions, as shown in Figure
\ref{fig:model_fit}. Given the deviations between the models and data,
caution is warranted in using the physical parameters listed in
Table \ref{table:model}, however some general trends do emerge.  First,
we examined the relative effect of the distance restriction.  The
distance restricted fits prefer a cooler {\teff}, $\sim$ 500 K instead
of 700 K.  The
distance restricted fits also suggest a higher surface gravity and
older age, but there is no clear behavior in the mass determination.
We also note that most of the fits with unrestricted distances prefer
distances much larger than the measured parallax.

We adopt the restricted distance and wavelength sets ($J_r, H_r$
and $K_r$) for further discussion.  These clipped wavelength sets were
chosen to sample strong molecular absorption, mostly due to {\meth}.
Agreement between the data and model within these windows indicates
that the BT-Settl models may have the proper opacity included in these windows, but due to the
dearth of benchmark brown dwarfs \citep[][and references therein]{2010ApJ...711.1087K}, the physical parameters
derived from these fits may have large systematic uncertainties.
These fits suggest an object with {\teff} $\simeq$ $500-600$ K, {\logg} $\simeq$
$4.2-5.0$, mass $\sim 10-30$ {\mjup} and an age of $1-5$ Gyr.  These
parameters agree with the results of \cite{2010MNRAS.408L..56L}, who
reported {\teff}  $ = 480-560$ K,  {\logg} $ = 4.0-4.5$, mass
$= 5-15$ {\mjup} and an age of $0.2-2.0$ Gyr.    
To examine the accuracy in other bandpasses, synthetic photometry was
computed in the IRAC and WISE bands reported in Table
\ref{table:properties} using the {\teff} $= 500$ K, {\logg}
= 4.0 model and reported parallax.  In general, the synthetic
photometry matched the reported values within $<$ 1 mag, and agreed
within $< 0.05$ mag for the $W1$ and $W3$ bandpasses, suggesting the
calculated opacities in these filters may be correct.  We note that the $H_r$
region produced identical results with and without the distance
restriction, suggesting that the BT-Settl models may perform well in this
wavelength range.  However, adopting the uncertainties from the physical
parameters from only this range probably underestimates the systematic
errors, especially since only
one set of models was considered.

\begin{center}
\begin{deluxetable*}{lllllllll}
\tablewidth{0pt}
 \tablecaption{Atmosphere Model Measurements}
 \tabletypesize{\scriptsize}

 \tablehead{
 \colhead{Parameter} & 
 \colhead{Full Spectrum} & 
 \colhead{$y$} &
 \colhead{ $J$} & 
 \colhead{$J_r$} & 
 \colhead{$H$} & 
 \colhead{$H_r$} & 
 \colhead{$K$} & 
 \colhead{$K_r$}
}
\startdata
Fit Ranges ($\micron$) & 0.9--2.35 & 0.9--1.15 & 1.15--1.35 & 1.27--1.31 & 1.45--1.8 & 1.50--1.57, & 1.9--2.35 & 2.1--2.18 \\
&  &  &  &  &  &  1.63--1.75 & & \\
DOF & 3409 & 982 & 743 & 182 & 1016 & 585 & 659 & 220 \\
\cline{1-9}
\multicolumn{9}{c}{No Distance Restriction} \\
\cline{1-9}
Min $\chi^2$ & 87.4 & 19.0 & 66.3 & 62.7 & 107 & 35.4 & 10.9 & 7.5 \\
{\teff} (K) & 700$\pm$50 & 750$\pm$70 & 600$\pm$50 & 620$\pm$60 & 590$\pm$60 & 500$\pm$50 & 610$\pm$60 & 630$\pm$70 \\
{\logg} (cgs) & 4.0$\pm$0.3 & 3.5$\pm$0.3 & 4.5$\pm$0.3 & 4.5$\pm$0.3 & 4.1$\pm$0.4 & 4.4$\pm$0.3 & 3.7$\pm$0.4 & 3.8$\pm$0.4 \\
Mass ({\cal M}$_{\sun}$) & 0.005 & 0.003 & 0.012 & 0.012 & 0.008 & 0.011 & 0.004 & 0.004 \\
Age (Gyr) & 0.08 & 0.02 & 0.9 & 0.9 & 0.5 & 1.5 & 0.07 & 0.09 \\
d (pc) & 49 & 61 & 22 & 20 & 27 & 6.4 & 23 & 24 \\
\cline{1-9}
\multicolumn{9}{c}{Distance Restriction} \\
\cline{1-9}
Min $\chi^2$ & 118 & 74.3 & 196 & 90.6 & 118 & 35.4 & 20.9 & 12.1 \\
{\teff} (K) & 500$\pm$50 & 500$\pm$50 & 500$\pm$50 & 600$\pm$50 & 500$\pm$50 & 500$\pm$50 & 520$\pm$70 & 510$\pm$60 \\
{\logg} (cgs) & 4.0$\pm$0.3 & 4.0$\pm$0.3 & 4.0$\pm$0.3 & 5.0$\pm$0.3 & 4.4$\pm$0.3 & 4.4$\pm$0.3 & 4.4$\pm$0.5 & 4.2$\pm$0.4 \\
Mass ({\cal M}$_{\sun}$) & 0.005 & 0.005 & 0.005 & 0.029 & 0.010 & 0.011 & 0.013 & 0.009 \\
Age (Gyr) & 0.2 & 0.2 & 0.3 & 5.1 & 1.3 & 1.4 & 2.1 & 1.0 \\
d (pc) & 4.6 & 4.2 & 4.9 & 6.7 & 6.2 & 6.4 & 4.2 & 3.5 \\
\enddata
 \label{table:model}
\end{deluxetable*}
\end{center}

\section{Conclusions}\label{sec:conclusions}
We have presented an analysis of FIRE observations of one of the
coldest brown dwarfs known, the T9 infrared spectral standard
{\namesh}\footnote{The FIRE spectrum is available online at \url{http://personal.psu.edu/jjb29/0722.html}.}.  Using current line lists and
atmospheric models, we characterized the NIR SED and constrained the
physical parameters of
{\namesh}.  At the resolutions achieved
with FIRE, we are able to identify individual molecular features
throughout the spectrum.  As more cold brown dwarfs are discovered through
new surveys such as WISE \citep{2010AJ....140.1868W}, the Canada--France Brown Dwarf Survey
 \citep[CFBDS;][]{2011AJ....141..203A} and VISTA \citep{2004SPIE.5493..411I},  these
features may be useful in determining fundamental parameters and
discriminating between spectral types.  Unfortunately, the BT-Settl model
atmospheres do not adequately reproduce the spectral features observed
at these low temperatures across the NIR regime, although fits over
restricted regions are more robust.  By limiting our analysis to the
wavelength and distance restricted fits, we derive a {\teff} of
$500-600$ K, {\logg} of $4.2-5.0$, mass of $10-30$ {\mjup} and age of $1-5$
Gyr.  These values agree well with the results from
\cite{2010MNRAS.408L..56L}, however the data should be re--examined as
models are further refined.

The radial velocity of {\namesh}  was measured as 46.9 {\kms} to a precision of
a few {\kms}.  Combined with parallax and proper motion measurements
from \cite{2010MNRAS.408L..56L}, the Galactic orbit of {\namesh} was
computed to investigate its parent population.   Its orbit is similar to many thin
disk objects, exhibiting low eccentricity and vertical excursions
taking it only $\sim$ 60 pc away from the plane.   This orbit also
agrees well with the age of $1-5$ Gyr derived from the atmospheric
fits.   As larger catalogs
of cold brown dwarfs with well measured kinematic properties are
assembled, their ensemble properties will be important for testing the
predictions of Milky Way kinematic structure models \citep[i.e.][]{2008ApJ...684L..79R,2011ApJ...737....8L}.

The rotational velocity of {\namesh} was also measured, employing the
BT-Settl atmospheric model as a template.  The object is
rotating at $\sim$ 40 $\pm$ 10 {\kms}, similar to other late--type L
and T dwarfs and further evidence that rotational braking is not
efficient in brown dwarfs \citep{2006ApJ...647.1405Z,2008ApJ...684.1390R}. 
In future investigations, we will secure observations of
brown dwarfs with small projected rotational velocities, such as
SDSS$~$J134646.45$~-$003150.4 \citep{2006ApJ...647.1405Z} to serve as
empirical templates.

\acknowledgements

We would like to thank the entire Magellan and Las Campanas staff for
their support and guidance during the assembly, installation and
commissioning of FIRE.  We would especially like to thank our
telescope operator during commissioning, Mauricio Martinez.  FIRE was constructed with support from the
NSF/MRI grant AST-0649190 and Curtis Marble.  JJB acknowledges the
financial support of NSF grant AST-0544588.  JJB thanks
Sergey Yurchenko and Bob Barber for assistance with line lists and Robyn
Sanderson for enlightening conversations on Galactic orbits.  RAS
acknowledges the financial and corporal support of the AJB chair.  We
thank the anonymous referee for their comments which greatly improved
the clarity and content of this manuscript.

{\it Facilities:} \facility{Magellan}

\end{document}